\documentclass[aps,prc,a4paper,nofootinbib,
preprintnumbers,superscriptaddress,showpacs,showkeys,twocolumn]{revtex4}

\usepackage{amsmath}
\usepackage{amssymb}
\usepackage{bm}
\usepackage{graphicx}
\usepackage{color}

\newcommand{\be}{\begin{equation}}
\newcommand{\ee}{\end{equation}}
\newcommand{\ba}{\begin{eqnarray}}
\newcommand{\ea}{\end{eqnarray}}

\begin{document}

\title{The cathode tube effect: heavy quarks probing the Glasma in p-Pb collisions}

\author{Marco Ruggieri}\email{ruggieri@lzu.edu.cn}
\affiliation{School of Nuclear Science and Technology, Lanzhou University, 222 South Tianshui Road, Lanzhou 730000, China.}

\author{Santosh K. Das}
\affiliation{School of Nuclear Science and Technology, Lanzhou University, 222 South Tianshui Road, Lanzhou 730000, China.}


\begin{abstract}
We study the propagation of charm quarks in the early stage of high energy proton-lead collision, considering the
interaction of these quarks with the evolving Glasma by means of the Wong equations. 
Neglecting quantum fluctuations at the initial time
the Glasma is made of longitudinal fields, but the dynamics leads to a quick formation of transverse fields;
we estimate such a formation time as $\Delta t\approx 0.1$ fm/c which is of the same order of the formation time
of heavy quark pairs $t_\mathrm{formation}\approx 1/(2m)$.
Limiting ourselves to the simple case of a static longitudinal geometry,
we find that heavy quarks are accelerated by the strong transverse color fields in the early stage
and this leads to a tilting of the $c-$quarks spectrum towards higher $p_T$ states.
This average acceleration can be understood in terms 
of drag and diffusion of $c-$quarks in a hot medium and
appears to be similar to the one felt by the electrons ejected by the electron cannon in a cathode tube:
we dub this effect as {\it cathode tube effect}.
The tilting of the spectrum affects the nuclear modification factor, 
$R_\mathrm{pPb}$, suppressing this below one at low $p_T$ and making it larger than one at intermediate $p_T$. 
We compute $R_\mathrm{pPb}(p_T)$ after the evolution of charm quarks in the gluon fields and we 
find that its shape is in qualitative agreement with the measurements of the same quantity 
for $D-$mesons in proton-lead collisions.  
\end{abstract}

\pacs{12.38.Aw,12.38.Mh}

\keywords{Relativistic heavy ion collisions, evolving Glasma, classical Yang-Mills fields, heavy quarks, 
nuclear modification factor, quark-gluon plasma}

\maketitle


{\em Introduction.} The study of the initial condition of the system produced by high energy collisions  
is a difficult but interesting problems related to the physics of
relativistic heavy ion collisions (RHICs), as well as to that of high energy proton-proton (pp) and 
proton-nucleus (pA) collisions. If the energy of the collision is very large then the two colliding nuclei
in the backward light cone can be described within the color-glass-condensate (CGC) effective theory 
\cite{McLerran:1993ni,McLerran:1993ka,McLerran:1994vd,Gelis:2010nm,Iancu:2003xm,McLerran:2008es,Gelis:2012ri},
in which fast partons are frozen by time dilatation and act as static sources
for low momentum gluons: their large occupation number allows for a classical treatment of these fields.  
The collision of two colored glass sheets, each representing one of the 
colliding objects in high energy collisions, leads to the formation of strong gluon fields in the forward light cone
named as the Glasma
\cite{Kovner:1995ja,Kovner:1995ts,Gyulassy:1997vt,Lappi:2006fp,Fukushima:2006ax,
Fries:2006pv,Chen:2015wia,Fujii:2008km,Krasnitz:2000gz,Krasnitz:2003jw,Krasnitz:2001qu}.
In the weak coupling regime the Glasma consists of longitudinal color-electric and color-magnetic fields;
these are again characterized by large gluon occupation number,
$A_\mu^a \simeq 1/g$ with $g$ the QCD coupling, 
so they can be described by classical field theory namely the Classical Yang-Mills  (CYM) theory.
Finite coupling bring up quantum fluctuations on the top of the Glasma \cite{Romatschke:2005pm,Romatschke:2006nk,Fukushima:2011nq,
Fukushima:2013dma,Iida:2014wea,Gelis:2013rba,Epelbaum:2013waa,Tanji:2011di,Ryblewski:2013eja,Ruggieri:2015yea,
Berges:2012cj,Berges:2013fga,Berges:2013lsa,Berges:2013eia,Ruggieri:2017ioa} that we 
do not consider in the present letter leaving their inclusion to a forthcoming study.  
Among the high energy collisions mentioned above, pA 
are interesting because they allow for both a theoretical
and an experimental study of the cold nuclear matter effects (CNME), namely those effects that are not directly
related to the formation of the QGP and that include shadowing \cite{Eskola:2009uj} as well as gluon 
saturation \cite{Fujii:2013yja,Ducloue:2015gfa,Rezaeian:2012ye}, 
see \cite{Albacete:2013ei,Albacete:2016veq,Prino:2016cni,Andronic:2015wma} for reviews.

Heavy quarks are excellent probes of the system created in high energy nuclear collisions,
both for the pre-equilibrium part and for the thermalized quark-gluon plasma (QGP), see
\cite{Rapp:2018qla, Aarts:2016hap, Greco:2017rro,Das:2015ana, Das:2013kea,Das:2016cwd,
Das:2017dsh,Das:2015aga,Beraudo:2015wsd,Xu:2015iha,Ozvenchuk:2017ojj,
Prino:2016cni,Andronic:2015wma,Mrowczynski:2017kso} and references therein.
Their formation time is very small in comparison with the one of light quarks:
indeed, this can be estimated as  $\tau_\mathrm{form} \approx 1/(2m)$ with
$m$ the quark mass which gives $\tau_\mathrm{form} \leq 0.1$ fm/c for the charm quark. 
Because heavy quarks are produced
immediately after the collision, they can propagate in the evolving Glasma fields and probe its evolution.
For nucleus-nucleus collisions it is likely that the effect of this propagation is largely washed out by the successive interaction
with the bulk quark-gluon plasma (QGP); on the other hand, for pA and pp collisions the effect of the interaction
with a medium is much smaller because of the smaller lifetime of the latter (if a QGP is created at all), therefore some effect of the initial
propagation in the gluon fields might survive up to the final stage of the evolution after hadronization. 
Moreover, heavy quark propagation in the background of the gluon field will hardly affect the latter,
due to the large mass and the small number of these quarks which leads to a negligible color current.
Therefore, heavy quarks are ideal probes of the strong gluon fields formed in high energy nuclear collisions.

In this study we focus on $c-$quarks in high energy p-Pb collisions. 
The main purpose of our study is to
compute consistenly the propagation of the heavy quarks in the initial gluon fields,
assuming besides that a QGP is not formed
or at least that it is formed at a later stage and its lifetime is quite smaller than the one in Pb-Pb collisions.
In particular, we are interested to the nuclear modification factor, $R_{\mathrm{pPb}}$,
for $D-$mesons that has
been measured recently~\cite{Abelev:2014hha, Aaij:2017gcy}. In fact, 
we find that the propagation in gluon fields leads to $R_{\mathrm{pPb}}$
that reminds at least qualitatively the one measured for $D-$mesons in p-Pb collisions.

Propagation of heavy quarks in the Glasma has been studied previously in \cite{Mrowczynski:2017kso}
although within a simplified approach based on a Fokker-Planck equation. Despite studying a simplified
situation, the work in \cite{Mrowczynski:2017kso} is interesting because it shows how the evolution of
$c-$quarks in Glasma can be interpreted in terms of drag and diffusion in momentum space, similarly 
to the evolution in a thermal medium.
Here we aim to perform a more complete study of the same problem without relying on the small
transferred momentum expansion of \cite{Mrowczynski:2017kso}, as well as including the dynamical
evolution of the gluon medium that is missing in \cite{Mrowczynski:2017kso},
in order to quantify the effect of the evolution of $c-$quarks in the Glasma on observables.
This is achieved by solving consistently the classical equations of motion of the gluon fields,
namely the classical Yang-Mills equations, and of the heavy quarks propagating in the Glasma,
that are the Wong equations. This approach is equivalent to solve the Boltzmann-Vlasov equations
for the heavy quarks in a collisionless plasma: as a matter of fact, 
the Boltzmann-Vlasov equations can be solved by means of the test particles method
which amounts to solve the classical equations of motion of the test particles, here represented by the heavy quarks,
and these classical equations are just the Wong equations. 

The purpose of our study is twofold. Firstly, 
we aim to estimate the impact of the early stage of p-Pb collisions on $R_{\mathrm{pPb}}$.
Secondly, we notice that this effect does not come alone, in the sense that
in our calculation the modification to $R_{\mathrm{pPb}}$
comes entirely from the propagation in the evolving Glasma: as a consequence, the shape of $R_{\mathrm{pPb}}$
that we find (which qualitatively agrees with experimental data) can be understood as the signature 
that the Glasma leaves on this observable.

We need to mention that the present study should be considered as a preliminary one since we do not include
a longitudinal expansion in our calculation, therefore we do not attempt to a serious
comparison with the existing experimental data: while the inclusion of the expansion might reduce the effect
on $R_\mathrm{pPb}$, we find that the largest part of it comes within $\approx 1$ fm/c of evolution,
therefore  most likely at least part of this effect will remain also in case the longitudinal expansion is taken into account
(we will include the longitudinal expansion anyway in a forthcoming paper).
Keeping this in mind, whenever we mention that we consider p-Pb collisions at a given energy
it means that we have set up the initial color charge distributions on the proton and Pb sides in 
agreement with what should be done for simulations of realistic collisions,
trying to keep both the color charge distributions and the saturation scales as closer as possible to what
should be done in a complete calculation where expansion is taken into account.

{\em Glasma and classical Yang-Mills equations.} 
In this section we briefly review the Glasma and the McLerran-Venugopalan (MV) 
model \cite{McLerran:1993ni,McLerran:1993ka,McLerran:1994vd,Kovchegov:1996ty}. 
We remark that in our notation the gauge fields have been rescaled by the QCD coupling
$A_\mu \rightarrow A_\mu/g$.
In the MV model, the static color charge densities $\rho_a$   
on the nucleus $A$ are assumed to be random variables that are 
normally distributed with zero mean 
and variance specified by the equation
\begin{equation}
\langle \rho^a_A(\bm x_T)\rho^b_A(\bm y_T)\rangle = 
(g^2\mu_A)^2 \varphi_A(\bm x_T)\delta^{ab}\delta^{(2)}(\bm x_T-\bm y_T);
\label{eq:dfg}
\end{equation}
here $A$ corresponds to either the proton or the Pb nucleus, $a$ and $b$ denote the adjoint color index;
in this work we limit ourselves for simplicity to the case of the $SU(2)$ color group therefore
$a,b=1,2,3$.
In Eq.~(\ref{eq:dfg}) $g^2\mu_A$ denotes the color charge density and it is of the order of the saturation momentum $Q_s$ \cite{Lappi:2007ku}.

The function $\varphi_A(\bm x_T)$  in Eq.~\eqref{eq:dfg} allows for a nonuniform probability distribution
of the color charge in the transverse plane.
In this letter we study the gluon fields produced in p-Pb collisions.
For the case of a the Pb nucleus we assume a uniform 
probability and take $\varphi(\bm x_T)=1$.  
On the other hand, for the proton we use the constituent
quark model \cite{Schenke:2014zha,Schenke:2015aqa,Mantysaari:2017cni,Mantysaari:2016jaz}: 
for each event, we firstly extract the position of the three valence quarks, $\bm x_{i}$ with
$i=1,2,3$, assuming a gaussian distribution, namely
\begin{equation}
\psi(\bm x_T)=e^{-(\bm x_T^2)/(2 B_{cq})};\label{eq:xi_proton}
\end{equation} 
then we build up the probability density
\begin{equation}
\varphi_p(\bm x_T) = \frac{1}{3}\sum_{i=1}^3 e^{-(\bm x_T^2 -\bm x_{i}^2 )/(2 B_\rho)}.\label{eq:xi_distri}
\end{equation}
The two parameters in Eqs.~\eqref{eq:xi_proton} and~\eqref{eq:xi_distri} are
$B_{cq} = 3$ GeV and $B_q = 0.3$ GeV.
We remark that this procedure does not correspond to assume that the three valence quarks
are the only sources of the large $x$ color charges: indeed, 
from Eq.~\eqref{eq:dfg} it should be  obvious to any reader familiar with the MV model that we 
distribute sea color charges analogously to what is done for the case of a homogeneous $g^2\mu$.
In fact, the constituent quark models amounts simply to assume that 
the large $x$ charges from the sea localize around the valence quarks: these act as seeds for the sea charges.
The sensitivity on the number of constituent hot spots of color charge
has been studied in \cite{Mantysaari:2016jaz}; in \cite{Schenke:2014zha,Schenke:2015aqa,Mantysaari:2017cni,Mantysaari:2016jaz}
the significance of this model in comparison with the simpler gaussian one is well explained. 
The gaussian model of the proton can be used as well
in our study
and we will report on this in a future work.

For the proton $g^2\mu_p \varphi_p(\bm x_T)^{1/2}$ 
can be understood as an $\bm x_T-$dependent $g^2\mu$
because $\varphi_p(\bm x_T)$ localizes
the distribution around the valence quarks: we fix $g^2\mu_p$ 
for each event assuming that 
$\langle  g^2\mu_p \varphi_p(\bm x_T)^{1/2}\rangle/ Q_s= 0.57$ following the result of \cite{Lappi:2007ku},
where the average
is defined with $\varphi_p(\bm x_T)$ as a weight function,
then estimating 
$Q_s$ at the relevant energy by using 
the standard GBW fit  \cite{GolecBiernat:1999qd,GolecBiernat:1998js,Kovchegov:2012mbw}
\begin{equation}
Q_s^2 = Q_{s,0}^2 \left(\frac{x_0}{x}\right)^\lambda,\label{eq:QS_slide}
\end{equation}
with $\lambda=0.277$, $Q_0=1$ GeV and $x_0=4.1\times 10^{-5}$.
We remind that whenever we apply this equation to high energy collisions,
the relevant value of $x$ for the two colliding objects can be estimated at midrapidity as $\langle p_T\rangle/\sqrt{s}$
where $\langle p_T\rangle$ corresponds to the average $p_T$ of the gluons produced
by the collision.
For example, at the RHIC energy for $x=0.01$ 
we obtain $Q_s = 0.47$ GeV in agreement with the estimate of \cite{Albacete:2012xq}.
At the LHC energy $\sqrt{s}=5.02$ TeV we find $Q_s = 0.80$ GeV which gives
$\langle  g^2\mu_p \varphi_p(\bm x_T)^{1/2}\rangle = 1.41$ GeV.

For the Pb nucleus the uncertainty on the $Q_s$ as well as on $g^2\mu$ comes from
the different model used to compute $Q_s$ for a large nucleus. Indeed the GBW fit in this case is modified as
\begin{equation}
Q_s^2 = f(A)Q_{s,0}^2 \left(\frac{x_0}{x}\right)^\lambda,\label{eq:QS_slide2}
\end{equation}
where
\begin{equation}
f(A) = A^{1/3}
\end{equation}
within a naive scaling hypothesis , and
\begin{equation}
f(A) =c A^{1/3}\log A
\end{equation}
within the IP-Sat model \cite{Kowalski:2007rw}. 
While other forms of $f(A)$ are possible  \cite{Armesto:2004ud,Freund:2002ux}, the two above give the higher and lower value of $Q_s$ at
the RHIC energy \cite{Lappi:2007ku} therefore we take these two to set the upper and lower estimate of $Q_s$.
Using again $Q_s/g^2\mu=0.57$ we find 
$g^2\mu_\mathrm{Pb} = 2$ GeV and $g^2\mu_\mathrm{Pb} = 3$ GeV at the RHIC energy taking respectively the IP-Sat and
naive forms; the modified GBW fit then leads to  $g^2\mu_\mathrm{Pb} = 3.4$ GeV and $g^2\mu_\mathrm{Pb} = 5.2$ GeV
for the two cases at the LHC energy.

The static color sources $\{\rho\}$ generate pure gauge fields outside and on the light cone, which in the forward light cone combine 
and give the initial Glasma fields. In order to determine these fields
we  firstly solve the Poisson equations for the gauge potentials
generated by the color charge distributions of the nuclei $A$ and $B$, namely
\begin{equation}
-\partial_\perp^2 \Lambda^{(A)}(\bm x_T) = \rho^{(A)}(\bm x_T)
\end{equation}
(a similar equation holds for the distribution belonging to $B$). Wilson lines are computed as
$
V^\dagger(\bm x_T) = e^{i \Lambda^{(A)}(\bm x_T)}$, 
$W^\dagger(\bm x_T) = e^{i \Lambda^{(B)}(\bm x_T)}$,
and the pure gauge fields of the two colliding nuclei are given by
$
\alpha_i^{(A)} = i V \partial_i V^\dagger$,
$\alpha_i^{(B)} = i W \partial_i W^\dagger$.
In terms of these fields the solution of the CYM in the forward light cone
at initial time, namely the Glasma gauge potential, 
can be written as 
$A_i = \alpha_i^{(A)} + \alpha_i^{(B)}$~ for $i=x,y$ and $A_z = 0$,
and the initial longitudinal Glasma fields are \cite{Kovner:1995ja,Kovner:1995ts} 
\begin{eqnarray}
&& E^z = i\sum_{i=x,y}\left[\alpha_i^{(B)},\alpha_i^{(A)}\right], \label{eq:f1}\\
&& B^z = i\left(
\left[\alpha_x^{(B)},\alpha_y^{(A)}\right]  + \left[\alpha_x^{(A)},\alpha_y^{(B)}\right]  
\right),\label{eq:f2}
\end{eqnarray}
while the transverse fields are vanishing. 
It has been suggested that the gauge potentials should be computed by defining the Wilson lines
as path ordered exponentials of multiple layers of color charges in order to describe the propagation
of a colored probe through a thick nucleus \cite{Lappi:2007ku}:
we have checked that using multiple layers instead of a single layer of charge does not affect
considerably our results, and for the sake of simplicity we report here only the results obtained using
one single layer, leaving a more complete report to a forthcoming article.

The dynamical evolution that we study here is given by the 
classical Yang-Mills (CYM) equations. 
In this study we follow \cite{Iida:2014wea} therefore we refer
to that reference for more details.
The hamiltonian density is given by
\begin{equation}
H = \frac{1}{2}\sum_{a,i}E_i^a(x)^2 + \frac{1}{4}\sum_{a,i,j}F_{ij}^a(x)^2,
\label{eq:H} 
\end{equation}
where the magnetic part of the field strength tensor is  
\begin{equation}
F_{ij}^a(x) = \partial_i A_j^a(x) - \partial_j A_i^a(x)  + \sum_{b,c}f^{abc} A_i^b(x) A_j^c(x);
\label{eq:Fij}
\end{equation}
here $f^{abc} = \varepsilon^{abc}$ with $\varepsilon^{123} = +1$.
The equations of motion for the fields and conjugate momenta, namely the CYM equations, are 
\begin{eqnarray}
\frac{dA_i^a(x)}{dt} &=& E_i^a(x),\\
\frac{dE_i^a(x)}{dt} &=& \sum_j \partial_j F_{ji}^a(x) + 
\sum_{b,c,j} f^{abc} A_j^b(x)  F_{ji}^c(x).\label{eq:CYM_el}
\end{eqnarray}
We solve the above equations on a static box in three spatial dimension as in \cite{Iida:2014wea,Ruggieri:2017ioa}.

{\em Heavy quarks in the evolving Glasma.}
At the initial time we assume that the momentum distribution of $c-$quarks is the prompt one obtained
within 
Fixed Order + Next-to-Leading Log (FONLL) QCD which 
describes the D-mesons spectra in $pp$ collisions after fragmentation~\cite{FONLL, Cacciari:2012ny,Cacciari:2015fta} 
\begin{equation}
\left.\frac{dN}{d^2 p_T}\right|_\mathrm{prompt} = \frac{x_0}{(x_1 + p_T)^{x_2}};\label{eq:HQ_1}
\end{equation}
the parameters that we use in the calculations are $x_0=6.37\times 10^8$, $x_1=9.0$ and $x_2=10.279$.
Normalization of the spectrum is not relevant in this letter because we are interested to the nuclear modification factor
which is a ratio of the final over initial spectrum and this is unaffected by the overall normalization since the number of
heavy quarks is conserved during the evolution; the slope of the spectrum has been calibrated to a collision at 5.02 TeV.
Moreover, we assume that the initial longitudinal momentum vanishes
(in a longitudinally expanding geometry this condition can be replaced by the
standard Bjorken flow $y=\eta$).
Initialization in coordinate space is done as follows:
the tranverse coordinates distribution is built up by means of the
function $\psi(\bm x_T)$ in Eq.~\eqref{eq:xi_proton}, because we expect the
heavy quarks to be produced in the overlap region of proton and Pb nucleus that
coincides with the transverse area of the proton;
on the other hand, we use a uniform distribution for the longitudinal coordinate
(in a longitudinally expanding geometry this condition can be replaced by a uniform distribution in
spacetime rapidity).

The dynamics of heavy quarks in the evolving Glasma is studied by the Wong equations  \cite{Wong:1970fu,Boozer}, that
for a single quark can be written as
\begin{eqnarray}
&&\frac{d x_i}{dt} = \frac{p_i}{E},\\
&&E\frac{d p_i}{dt} = Q_a F^a_{i\nu}p^\nu,\\
&&E\frac{d Q_a}{dt} = - Q_c\varepsilon^{cba} \bm A_b\cdot\bm p;
\end{eqnarray}
where $i=x,y,z$; here, the first two equations are the familiar Hamilton equations of motion for the coordinate and its conjugate
momentum, while  the third equation corresponds to the gauge invariant color current conservation.
Here $E = \sqrt{\bm p^2 + m^2}$ with $m=1.5$ GeV corresponding to the charm quark mass. 
In the third Wong equation $Q_a$ corresponds to the
$c-$quarks color charge: we initialize this by a uniform distribution with support in the range $(-1,+1)$.
For each $c$ quark we produce a $\bar c$ quark as well: for this we assume the same
initial position of the companion $c$, opposite momentum and opposite color charge. 
Solving the Wong equations is equivalent to solve the Boltzmann-Vlasov equations for a collisionless plasma
made of heavy quarks, which propagate in the evolving Glasma;
in fact, the latter equation can be solved by means of the test particle method which amounts
to solve the classical equations of motion of the particles in the background of the evolving gluon field.
In principle, we should include the heavy quarks color current density
on the right hand side of Eq.~\eqref{eq:CYM_el} and compute the backreaction on the
gluon fields. However, we neglect this backreaction: 
this approximation is usually used to study the propagation of heavy probes in a thermal QGP bath
and sounds quite reasonable due to 
the small number of heavy quarks produced by the collision, as well as to their large mass, 
both of these factors leading eventually to a negligible color current density. 
On the transverse lattice we do not assume periodic boundary conditions for the heavy quarks:
as soon as a heavy quark reaches the boundary of the transverse box we cancel any interaction 
with the gluon fields and its motion becomes a simple free streaming.


\begin{figure}[t!]
\begin{center}
\includegraphics[scale=0.3]{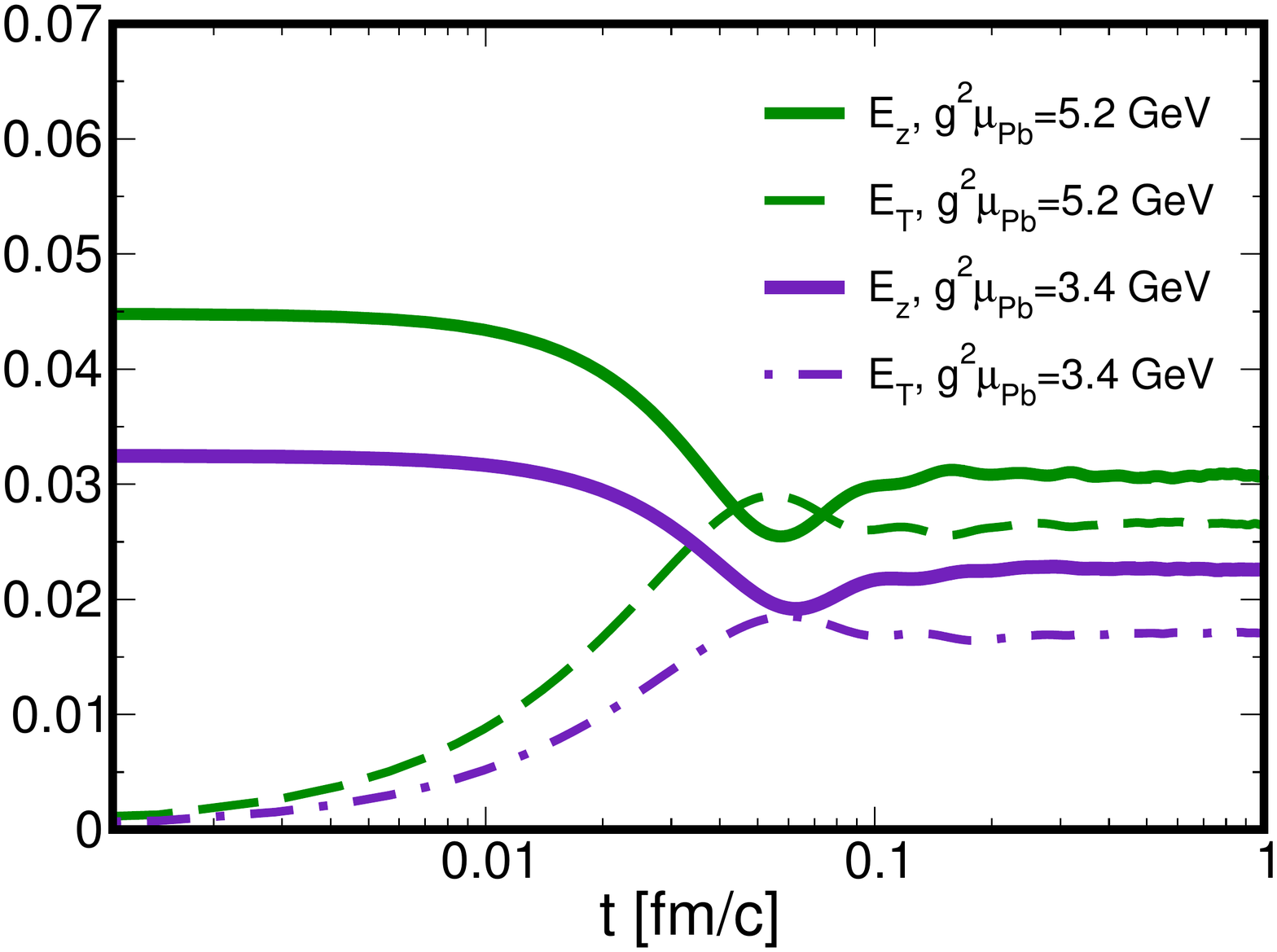}
\end{center}
\caption{\label{Fig:2}
Color online. Averaged color-electric fields for p-Pb collision, measured in lattice units.
Solid lines correspond to the longitudinal fields while dashed lines denote the transverse fields;
green and indigo lines correspond to $g^2\mu_\mathrm{Pb}=5.2$ GeV and 
$g^2\mu_\mathrm{Pb}=3.4$ respectively. Lattice spacing is $\delta x=0.04$ fm.}
\end{figure}

{\em Results.} In Fig.~\ref{Fig:2} we plot 
the averaged color-electric fields, measured in lattice units, versus time.
Solid lines correspond to the longitudinal fields while dashed lines denote the transverse fields;
green and indigo lines correspond to $g^2\mu_\mathrm{Pb}=5.2$ GeV and 
$g^2\mu_\mathrm{Pb}=3.4$ respectively.
The transverse size of the box is $4$ fm and we have used a transverse lattice with size $91\times 91$ that gives 
the lattice spacing $\delta x = 0.04$ fm.
At initial time the system is made of purely longitudinal fields, but 
this configuration is intrinsically unstable and the gluon dynamics leads to the production of
transverse fields: within $\Delta t \approx 0.1$ fm/c the bulk is already formed,
and at later times the magnitude of the several components of the fields does not change considerably.

\begin{figure}[t!]
\begin{center}
\includegraphics[scale=0.3]{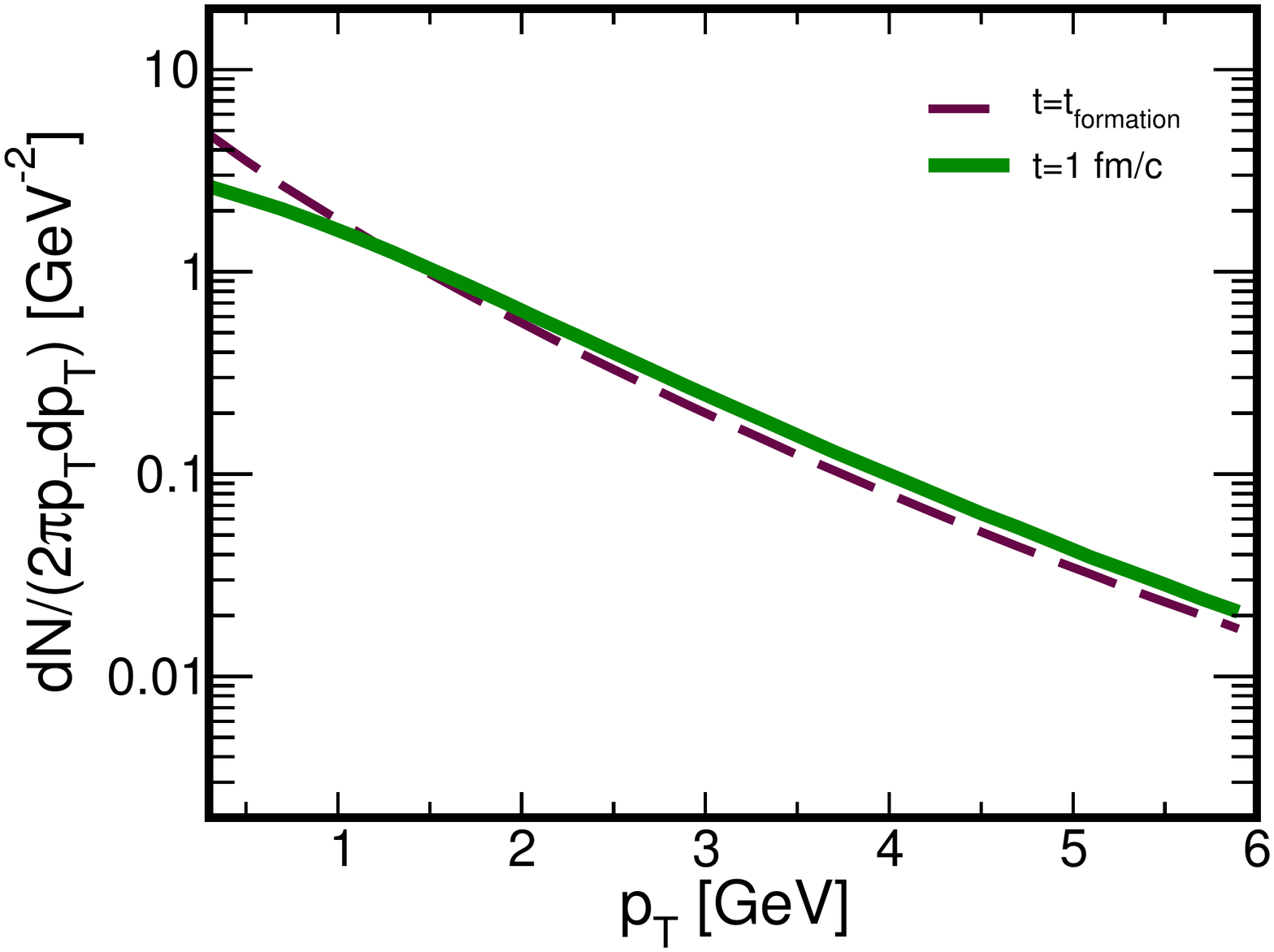}\\
\includegraphics[scale=0.3]{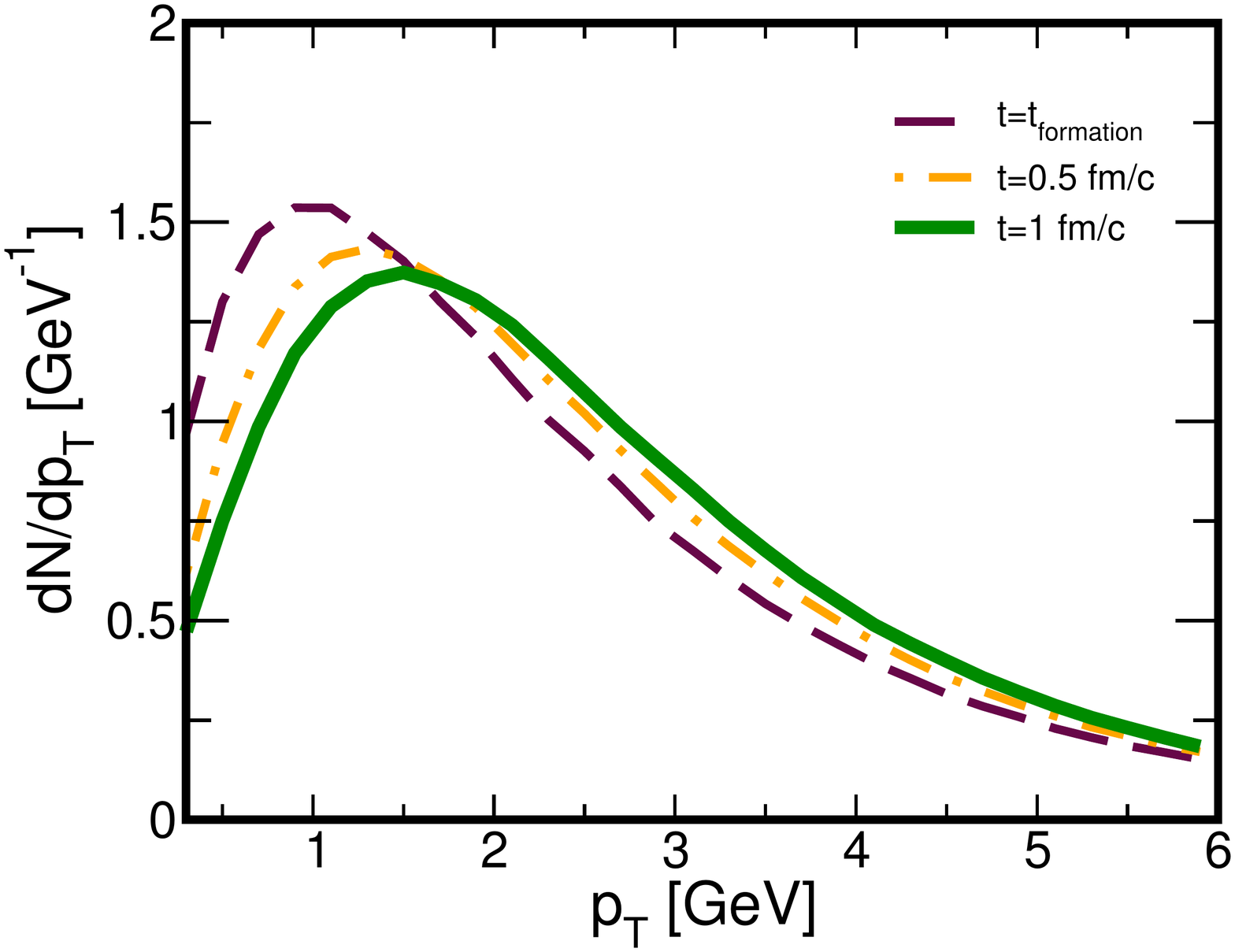}
\end{center}
\caption{\label{Fig:3}
Color online. In the upper panel we plot the D-meson spectrum, $dN/d^2 p_T$, at initial time (maroon dashed line)
and at $t=1$ fm/c (green solid line).
In the lower panel we plot the momentum distribution of $c-$quarks, $dN/d^2 p_T$,
at the initial time (dashed maroon line), at $t=0.5$ fm/c (orange dot-dashed line) and
at $t=1$ fm/c (green solid line). We take $g^2\mu_\mathrm{Pb} = 5.2$ GeV.}
\end{figure}

In the upper panel of Fig.~\ref{Fig:3} we plot the D-meson spectrum, $dN/d^2 p_T$, at initial time (maroon dashed line)
and at $t=1$ fm/c (green solid line).
In the lower panel of the same figure we plot the momentum distribution of $c-$quarks, $dN/d p_T$,
at the initial time (dashed maroon line), at $t=0.5$ fm/c (orange dot-dashed line) and
at $t=1$ fm/c (green solid line). 
We assume $g^2\mu_\mathrm{Pb} = 5.2$ GeV.
In the calculation we have assumed that the formation time
of $c-$quarks is $t_\mathrm{formation}=1/(2m_c)\approx 0.06$ fm/c for $m=1.5$ GeV
but we have checked that lowering this value does not affect considerably the final result.
At the end of the evolution  we adopt a standard fragmentation for the charm 
quark to D-meson~\cite{Pet}, with
\be
f(z) \propto 
\frac{1}{z \left( 1- \frac{1}{z}- \frac{\epsilon_c}{1-z} \right)^2}
\label{fg}
\ee
where $z=p_D/p_c$ is the momentum fraction of the D-meson fragmented from the charm quark and
$\epsilon_c$ is a free parameter to fix the shape of the fragmentation function in order to 
reproduce the D-meson production in $pp$ collisions~\cite{Scardina:2017ipo} namely $\epsilon_c=0.06$.
In the lower panel of Fig.~\ref{Fig:3} we plot the $c-$quark distribution $dN_c/dp_t$ at the initial time
(maroon dashed line), at $t=0.5$ fm/c (orange dot-dashed line) and at $t=1$ fm/c (green solid line).
We notice that the main effect of the interaction of the heavy quarks with the gluon field is to 
empty the low $p_T$ states of the $c-$quarks and fill the states with higher values of $p_T$: this effect looks 
similar to the acceleration that electric charges would feel in the background of a transverse field.

\begin{figure}[t!]
\begin{center}
\includegraphics[scale=0.3]{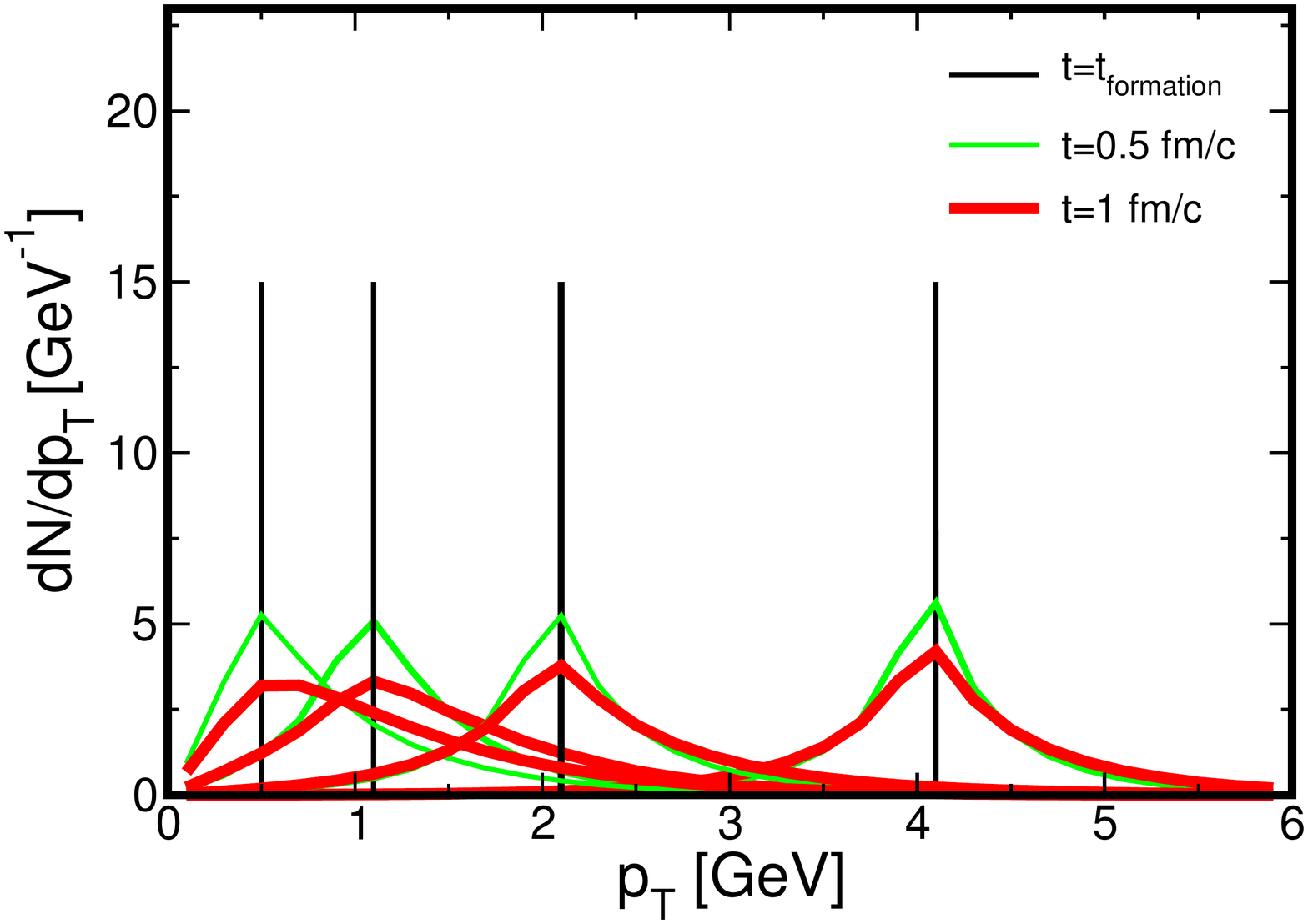}
\end{center}
\caption{\label{Fig:3a}
Color online. Evolution of $\delta-$distribution functions
of $c-$quarks in the Glasma fields. Black solid lines correspond to the initializations,
green dashed lines to $t=0.5$ fm/c and
red solid lines to $t=1$ fm/c. We take $g^2\mu_\mathrm{Pb} = 5.2$ GeV.}
\end{figure}

In order to understand better the interaction of the $c-$quarks with the evolving Glasma fields
we prepare initializations in which we put all the $c-$quarks in a very thin $p_T$ bin to obtain a $\delta-$like distribution;
the evolution of this distribution is studied again by means of the Wong equations.
This is done in order to better understand the interaction of the Glasma with different $p_T$ modes.
The results of this are shown in Fig.~\ref{Fig:3a} in which we plot the distribution function
$dN_c/dp_T$ at initial time (solid black lines), at $t=0.5$ fm/c (green dashed lines)
and at $t=1$ fm/c (solid red lines) for several values of the initial $p_T$.
We notice that in all the cases examined here the interaction with the Glasma fields
leads to the spreading of $dN/dp_T$, which is very similar to the standard diffusion in momentum space
encountered in a Brownian motion.
In addition to this, for low $p_T$ we find that diffusion is flanked by a drag towards higher values of 
$p_T$: this results in an average acceleration of the $c-$quarks
and it is similar to what we would expect putting low-$p_T$ quarks in a hot medium.
A more quantitative comparison of the evolution of heavy quarks in Glasma and in a hot plasma will be
the subject of a forthcoming article.

\begin{figure}[t!]
\begin{center}
\includegraphics[scale=0.3]{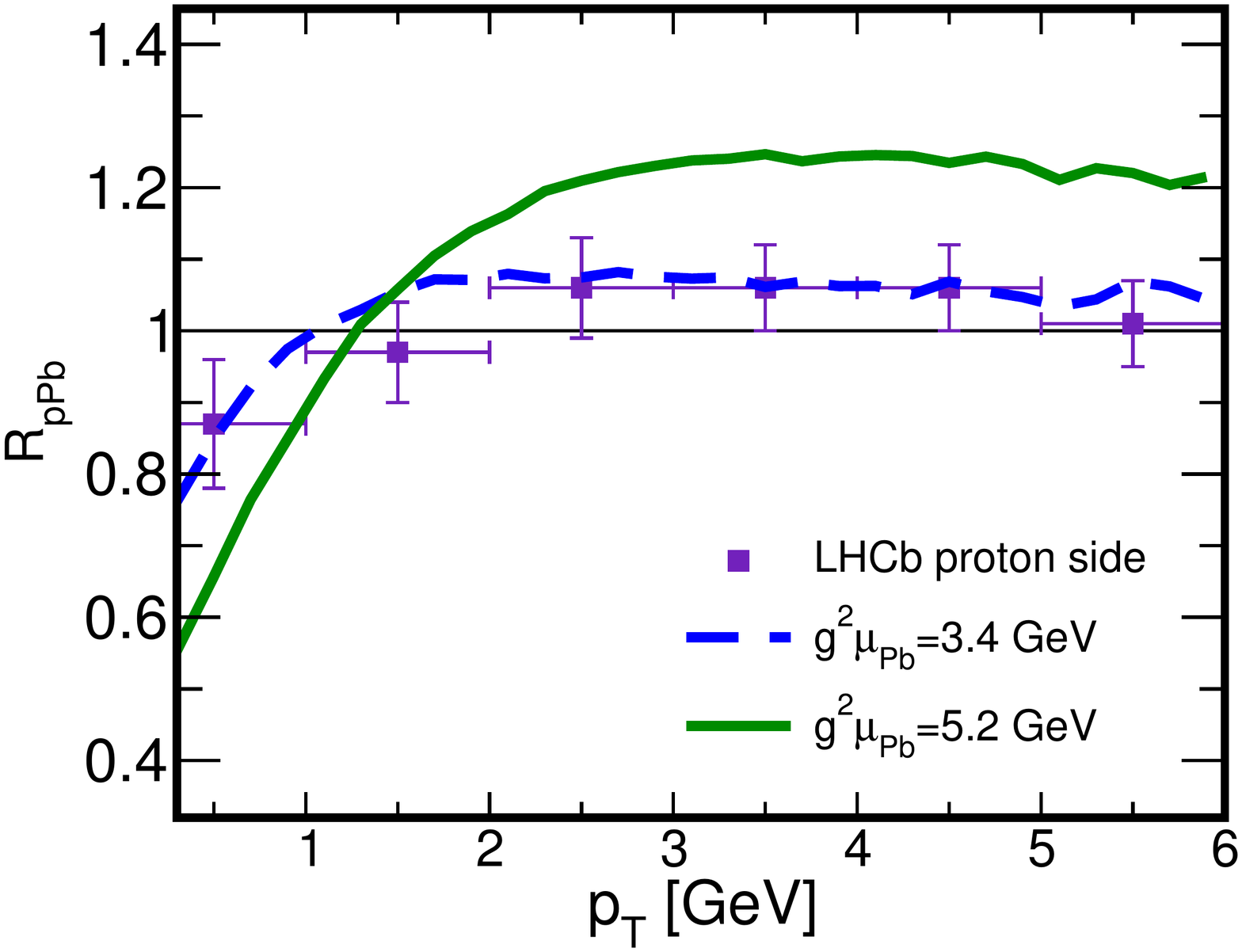}
\end{center}
\caption{\label{Fig:4}
Color online. Nuclear suppression factor versus $p_T$.
We plot the results for two values of $g^2\mu$ for the Pb nucleus, namely $g^2\mu = 3.4$ GeV
(orange dot-dashed line) and $g^2\mu = 5.2$ GeV (orange dashed line).
The solid green line corresponds to the $R_\mathrm{pPb}$ for the $D-$meson, obtained assuming a
standard fragmentation scenario for the $c-$quarks. Simulations have been stopped at $t=1$ fm/c.
Data correspond to the backward rapidity side (namely the proton side)
of the LHCb collaboration~\cite{Aaij:2017gcy}.}
\end{figure}

The drag and diffusion of the $c-$quarks in momentum space has an effect on the nuclear modification factor of $D-$meson, 
defined as
\begin{equation}
R_\mathrm{pPb} = \frac{\left(dN/d^2 p_T\right)_\mathrm{evolved}}{\left(dN/d^2 p_T\right)_\mathrm{prompt}},
\end{equation}
where the prompt spectrum is given by Eq.~\eqref{eq:HQ_1} after fragmentation and 
$\left(dN/d^2 p_T\right)_\mathrm{evolved}$ corresponds to the spectrum obtained by fragmentation
of the $c-$quark spectrum
after the evolution in the Glasma fields.
In Fig.~\ref{Fig:4} we plot the nuclear modification factor for the $D-$meson that we obtain within our calculation.
The result is shown for two values of $g^2\mu_\mathrm{Pb}$ for the Pb nucleus 
at $\sqrt{s}=5.02$ TeV, namely $g^2\mu_\mathrm{Pb} = 3.4$ GeV
(dashed blue line) and $g^2\mu_\mathrm{Pb} = 5.2$ GeV (solid green line)
as discussed in the previous section.
Experimental data correspond to the backward rapidity region (namely to the proton side)
obtained by the LHCb collaboration~\cite{Aaij:2017gcy}. We remark that although we show experimental data here,
we do not aim to a precise fit of these by our calculation because we miss the longitudinal expansion:
data are shown only to quantify the order of magnitude of our result, while a closer comparison with data
will be the subject of a forthcoming study. 
We have chosen to show these data rather than the averaged published by the ALICE collaboration
because those are an average of the forward and backward rapidity region,
and in this case  the CNME are very important and should be included in our initial state.
We have checked however that including these effects in the initial state does not affect the
drag and diffusion of $c-$quarks in the evolving Glasma (results will be reported eslewhere).

Figure~\ref{Fig:4} is the main result of the present letter: it shows that $R_\mathrm{pPb}$
can get a substantial deviation from one because of the interaction of the $c-$quarks with the evolving
gluon fields in the Glasma in the very early stage of a high energy p-Pb collision. 
As explained above, this result is due to the diffusion of heavy quarks in momentum space
accompanied by a drag of the low $p_T$ quarks towards higher momenta.
The net effect that we find is very different from what is usually discussed in the heavy quark community, namely energy loss. 
In fact, our results suggest that in the very early stage heavy quarks can gain energy
rather than loose it, because they are formed almost immediately after the collision and probe the strong gluon fields
of the Glasma while energy loss  will be substantial only in presence of a medium, namely of the quark-gluon
plasma that forms in a later stage. Most likely, this energy gain can be understood even in simpler terms considering that
low and intermediate $p_T$ heavy quarks are injected at the formation time into a system with a very large
energy density: therefore it appears natural that during their propagation they get energy rather than loose it.

This effect is interesting not only for its straightforward application to heavy quarks:
as a matter of fact, since it comes from the propagation in the strong gluon fields of the evolving Glasma,
the $c-$quarks probe these fields. The fact that the qualitative shape of our $R_\mathrm{pPb}$
resembles that measured in experiments might suggest that at least part of the measured $R_\mathrm{pPb}$
comes from the propagation of the $c-$quarks in the Glasma, and might be considered as the signature of the Glasma itself.
In this regards a
more quantitative statement will be put in a forthcoming article when the longitudinal expansion will be included,
and the amount of this effect will be compared to CNME.

We dub the effect summarized in Figs.~\ref{Fig:3} and~\ref{Fig:4} as the {\it cathode tube effect}.
The reason for this name is easy to understand. As a matter of fact, the cathode tubes are devices
in old televisions, in which an electron cannon ejects electrons and these are accelerated and deflected
by electric field before they hit a fluorescent screen. 
{\it Mutatis mutandis}, the same effect takes place in the early stage of high energy p-Pb collisions: 
indeed, here (color-)electric fields accelerate the prompt $c-$quarks that are injected into the bulk
by the inelastic collisions among the proton on the one hand and the nucleons in Pb on the other hand
(using this analogy, the electron cannon is here replaced by the p and Pb projectiles).

{\em Conclusions and outlook.}     
We have studied consistently the propagation of $c-$quarks in the evolving strong gluon fields
allegedly produced in high energy p-Pb collisions. As the initial condition we have taken the standard Glasma
with longitudinal color-electric and color-magnetic fields, adapted in order to take into account the finite size of the system;
for the initialization of the $c-$quarks we have considered the standard FONLL perturbative production
tuned in order to reproduce the $D-$meson spectrum in proton-proton collisions.
We have set up the saturation scale for both the proton and the Pb nucleus in order to reproduce the expected
one at $\sqrt{s}=5.02$ TeV: for this reason, even if we do not include the longitudinal expansion in the calculation,
we discuss about the gluon fields produced in p-Pb collisions at this energy.

We have computed the nuclear modification factor, $R_\mathrm{pPb}$, for these collisions: the result is summarized
in Fig.~\ref{Fig:4}. Although we do not aim to reproduce the experimental data because of the lack of the longitudinal expansion,
we have found that the qualitative shape of our $R_\mathrm{pPb}$ resembles that measured by the LHCb collaboration
on the proton side. Since in our calculation this shape comes directly from the propagation of the $c-$quarks in the
evolving Glasma fields, we suggest that at least part of the measured $R_\mathrm{pPb}$ is the signature
of the Glasma formed in high energy collisions. A firm statement will be put after we will have included the
longitudinal expansion in our calculation and this will be the subject of another article.
For the time being, we emphasize that the propagation of $c-$quarks in the evolving Glasma has only been 
partly studied within a small transferred momentum approximation 
and assuming a static gluonic medium \cite{Mrowczynski:2017kso},
so this letter aims to start to fill this gap and paves the way for more complete studies.

We remark that we have not assumed the formation of a hot medium, namely the QGP, in this calculation.
Indeed, although there is a lot of evidence that the QGP is formed in Pb-Pb collisions, such a strong evidence
is missing at the moment for p-Pb collisions. 
We will consider more closely this problem in the future, by coupling our evolution of the $c-$quark spectrum to
relativistic transport and to Langevin dynamics, in order to estimate quantitatively the effect of a hot medium on $R_\mathrm{pPb}$.

We have preliminarly studied the effect of the propagation of the $c-$quarks in the evolving Glasma 
in the case of Pb-Pb collisions at the LHC energy. In this case we have checked that
a propagation for approximately $0.3$ fm/c, which is a standard initialization time for QGP in relativistic transport and hydro simulations,
is enough to obtain a substantial effect. Although the $R_\mathrm{PbPb}$ in this case cannot be compared directly with the
experimental data due to the much longer propagation in the hot QGP, the effect of the early propagation in the
gluon fields should not be ignored. Again, we will couple our results to relativistic transport \cite{Plumari:2017ntm,Das:2015ana} 
in the near future in order to quantify
how the tilting of the $c-$quarks spectrum produced in the pre-equilibrium phase affects the late stage dynamics
of heavy quarks in Pb-Pb collisions.

We have not included for simplicity the effects of CNME on the prompt spectrum in our calculations.
Because of the lack of these effects in the present calculation, 
most likely the cathode tube effect is more relevant for the
proton side of the p-Pb collision in which shadowing and/or gluon saturation should not give a substantial contribution.
In fact, experimental data show that suppression of $R_\mathrm{pPb}$ is more pronounced
on the Pb side, where both shadowing and gluon saturation are expected to give substantial deviations from the
perturbative QCD prompt production of $c-$quarks. 
In addition to these, we have not included here the quantum fluctuations on the top of the Glasma:
these fluctuations add a transverse electric field at the initial time therefore they will enhanhce the cathode tube effect.
We will consider all these important effects in a forthcoming article.

\begin{acknowledgements}
M. R. acknowledegs discussions with the participants 
of the Next Frontiers in QCD 2018 Workshop held in Yukawa Institute for Theoretical Physics, Kyoto University,
where this work has been presented for the first time.
Comments from K. Fukushima and S. Mrowczynski have spurred the authors to study diffusion in more detail
and have lead to the results in Fig.~\ref{Fig:3a}.
Moreover, the authors acknowledge V. Greco for useful discussions during the preparation of this article.
The work of M. R. has been supported by the Fundamental Research Funds for the Central Universities.
\end{acknowledgements}

\end{document}